\documentstyle[prl,aps,floats,epsf]{revtex}
 
\begin{document}
 
\draft                    
 
\wideabs{
 
 
\title{
          Orientational Melting in Carbon Nanotube Ropes}
 
\author{ Young-Kyun Kwon
         and David Tom\'anek}
 
\address{Department of Physics and Astronomy, and
         Center for Fundamental Materials Research, \\
         Michigan State University,
         East Lansing, Michigan 48824-1116}
 
\date{Received 24 August 1999} 
 
\maketitle
 
 
\begin{abstract}
Using Monte Carlo simulations, we investigate the possibility of an
orientational melting transition within a ``rope'' of $(10,10)$ carbon
nanotubes. When twisting nanotubes bundle up during the synthesis,
orientational dislocations or twistons arise from the competition
between the anisotropic inter-tube interactions, which tend to align
neighboring tubes, and the torsion rigidity that tends to keep
individual tubes straight. We map the energetics of a rope containing
twistons onto a lattice gas model and find that the onset of a free
``diffusion'' of twistons, corresponding to orientational melting,
occurs at $T_{OM}{\agt}160$~K.
\end{abstract}
 
 
 
 
}
 
\narrowtext
 
Since their first successful synthesis in bulk quantity
\cite{Thess96}, ``ropes'' of single-wall carbon nanotubes have been in
the spotlight of nanotube research. Recent experimental data indicate
that carbon nanotube ropes exhibit an unusual temperature dependence
of conductivity \cite{Cond}, magnetoresistance \cite{Magnet}, and
thermoelectric power \cite{Therm}. Several physical phenomena have
been suggested to cause the intriguing temperature dependence of
conductivity behavior, such as twistons \cite{Kane-Mele},
orientational melting \cite{{PRLxNTR},{JMRxNTR}}, weak localization
\cite{weak-loc}, and Kondo effect \cite{Magnet}. The opening of a
pseudogap near $E_F$ in a bundle composed of (10,10) tubes has been
postulated to result from breaking the $D_{10h}$ tube symmetry by the
triangular lattice \cite{{PRLxNTR},{Louie98}}.
 
Nanotubes \cite{Iijima91} and C$_{60}$ ``buckyball'' molecules
\cite{{Smalley},{Huffman}} are similar inasfar as their interaction is
weakly attractive and nearly isotropic when condensing to a solid. The
small anisotropy of the C$_{60}$ intermolecular potential drives the
solid to an orientationally ordered simple-cubic lattice with four
molecules per unit cell at low temperatures \cite{Heiney91}. Only at
$T{\agt}249$~K does the C$_{60}$ solid undergo a transition to a
face-centered cubic lattice. As confirmed by $^{13}$C nuclear magnetic
resonance \cite{C60-rot}, this is an order-disorder phase transition,
with C$_{60}$ molecules spinning freely and thus becoming equivalent
above $249$~K. Unlike the C$_{60}$ solid, very little is known about
the equilibrium structure of bundled nanotubes beyond the fact that
they form a triangular lattice \cite{Thess96}. In particular, nothing
is known about the equilibrium orientation of the tubes within a rope.
More intriguing still is the possibility of an orientational melting
transition associated with the onset of orientational disorder within
a rope.
 
In the following, we calculate the potential energy surface and the
orientational order of straight and twisted tubes within a rope. We
further postulate that realistic nanotube ropes contain orientational
dislocations that have been frozen in as the tubes formed ropes at a
finite temperature, the same way as dislocations are known to form in
crystals. We map the energetics of a rope with dislocations onto a
lattice gas model and find that the onset of orientational disorder,
corresponding to a free axial diffusion of twistons, should occur at
$T_{OM}{\agt}160$~K.
 
In order to determine the orientational order and the rotational
motion of tubes in a rope, we need to describe both the inter-tube
interaction and the torsional strain within the individual tubes.
Since the anisotropic part of the inter-tube interaction in a rope is
weak and local, it can be well described by pairwise inter-tube ({\em
not} inter-atomic) interactions. To describe the interaction between
coaxial tubes as a function of their orientation, we use the
parametrized linear combination of atomic orbitals (LCAO) formalism
with parameters determined by {\em ab initio} results for simpler
structures \cite{PRLxCAR}. This technique has been successfully used
to explain superconductivity arising from inter-ball interactions in
the doped C$_{60}$ solid \cite{C60-supercond}, the opening of a
pseudogap near $E_F$ in a (10,10) nanotube rope
\cite{PRLxNTR,JMRxNTR}, and the opening and closing of four pseudogaps
during the librational motion of a (5,5)@(10,10) double-wall tube
\cite{PRLxDWT}. The total energy functional consists of a nonlocal
band structure energy term and of pairwise interatomic interactions
describing both the closed-shell repulsion and the long-range van der
Waals attraction. A smooth cutoff function \cite{cutoff} has been
implemented to keep the total energy continuous as the neighbor
topology changes while tubes rotate. This energy functional correctly
reproduces the exfoliation energy, the interlayer distance and the
$C_{33}$ modulus of hexagonal (AB) graphite, as well as the energy
barrier for interlayer sliding, corresponding to the energy difference
between AB and AA stacked graphite.
 
 
\begin{figure}
     \centerline{
         {\raisebox{0.37\columnwidth}{\large\bf (a)}}
         \hspace*{-0.1\columnwidth}
         \epsfxsize=0.37\columnwidth
         \epsffile{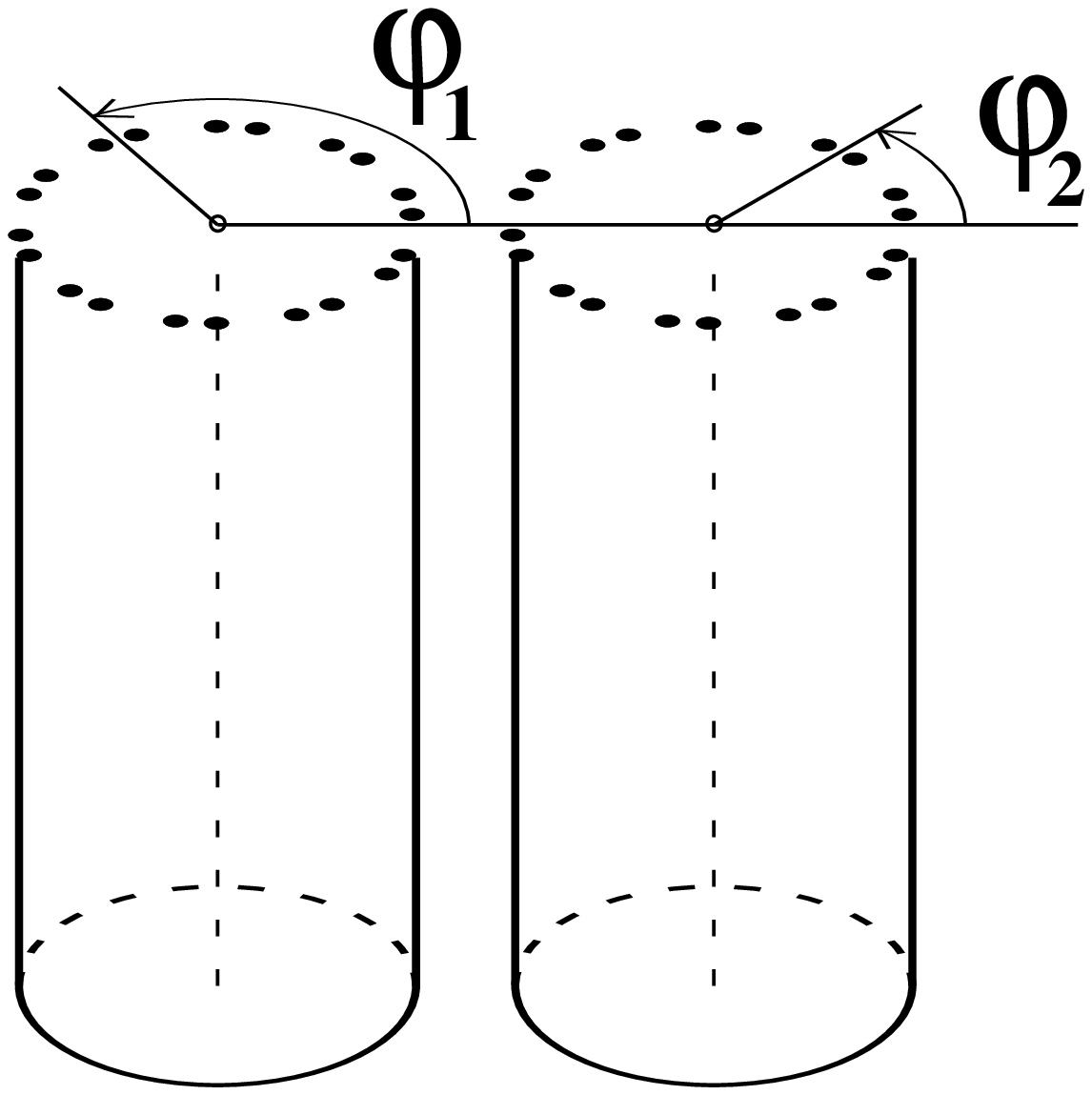}
         \hspace*{0.01\columnwidth}
         {\raisebox{0.37\columnwidth}{\large\bf (b)}}
         \hspace*{-0.10\columnwidth}
         \epsfxsize=0.55\columnwidth
         \epsffile{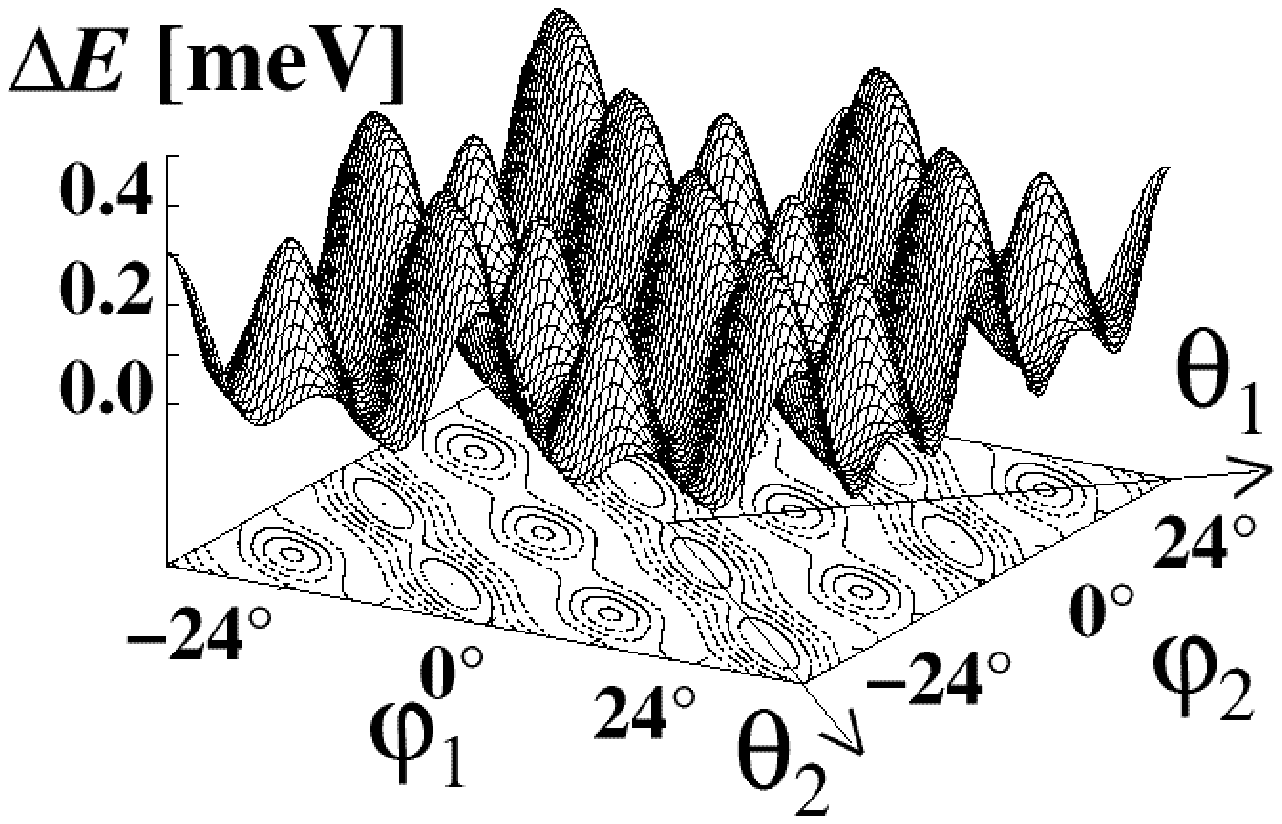}
     }
     \vspace*{0.01\columnwidth}
     \centerline{
         {\raisebox{0.30\columnwidth}{\large\bf (c)}}
         \hspace*{-0.15\columnwidth}
         \epsfxsize=0.55\columnwidth
         \epsffile{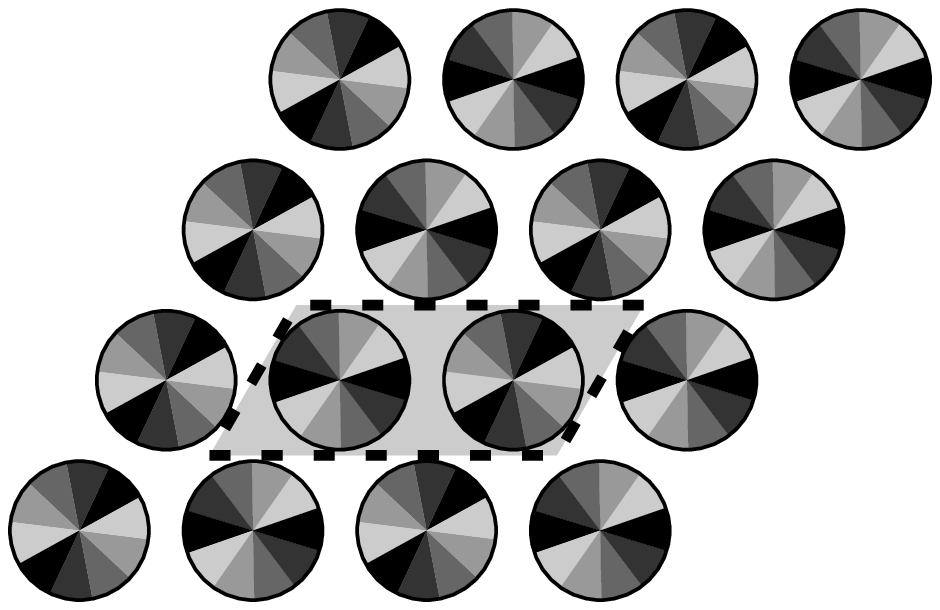}
         \hspace*{0.15\columnwidth}
         {\raisebox{0.27\columnwidth}{\large\bf (d)}}
         \hspace*{-0.2\columnwidth}
         \epsfxsize=0.35\columnwidth
         \epsffile{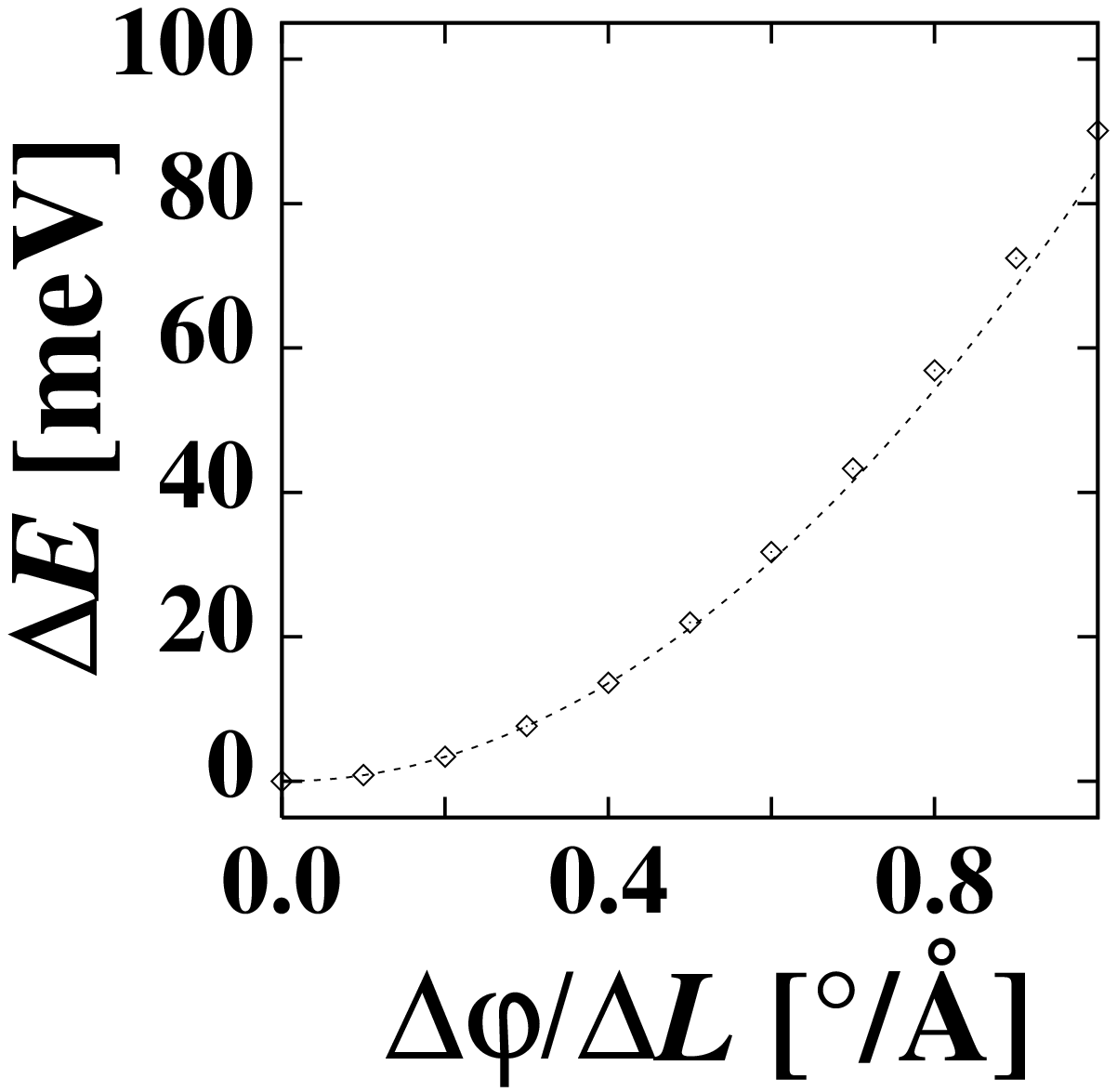}
    }
 \caption{
 (a) Definition of the orientational angles $\varphi_1$ and
 $\varphi_2$ of two aligned nanotubes.
 (b) Interaction energy {\em per atom} between two (10,10)
 carbon nanotubes at equilibrium distance as a function of the tube
 orientations.
 (c) Top view of the equilibrium structure of bundled interacting
 nanotubes, with a two-tube unit cell.
 (d) Torsion energy {\em per atom} within an individual (10,10)
 nanotube.
 }
 \label{Fig1}
 \end{figure}
 
To determine the interaction between a pair of aligned (10,10)
nanotubes, we first define the orientational angles $\varphi_1$ and
$\varphi_2$ for these tubes by the azimuthal angle of the center of a
particular bond with respect to the connection line between adjacent
nanotube axes, as shown in Fig.~\ref{Fig1}(a). For each
$(\varphi_1,\varphi_2)$ pair, we calculate the inter-tube interaction
using a fine mesh of 800 k-points sampling the one-dimensional
irreducible Brillouin zone. Due to the high symmetry of the system,
the interaction energy $\Delta{E}(\varphi_1,\varphi_2)$ is periodic in
$\varphi_1$ and $\varphi_2$, with a period $\Delta\varphi=36^\circ$.
With the simple variable transformation $\theta_1=\varphi_1+\varphi_2$
and $\theta_2=\varphi_1-\varphi_2$, the interaction energy $\Delta{E}$
can be simply expanded in harmonic functions of $\theta_1$ and
$\theta_2$, with the same period $\Delta\theta=36^\circ$. The
resulting potential energy surface, displayed in Fig.~\ref{Fig1}(b),
shows a maximum corrugation of only
$\Delta{E}_{max}{\approx}0.5$~meV/atom.
 
The equilibrium geometry of bundled (10,10) nanotubes can be found by
optimizing the total energy $E$ with respect to the orientations of
all individual tubes \cite{orientation}. Due to the high level of
orientational frustration in a triangular lattice of (10,10) nanotubes
with $D_{10h}$ symmetry, the potential energy surface
$E(\varphi_1,\varphi_2,\ldots)$ is very complex. We determine the
global minimum of $E(\varphi_1,\varphi_2,\ldots)$ by applying the
Metropolis Monte Carlo algorithm to an infinite system of straight
nanotubes in a $(2{\times}2)$, $(4{\times}4)$, $\ldots$ ,
$(8{\times}8)$ superlattice, with unit cells containing between 4 and
64 nanotubes. Independent of the unit cell size, we find that the
global energy minimum corresponds to a two-dimensional oblique lattice
with two tubes per unit cell, shown in Fig.~\ref{Fig1}(c). In
equilibrium, the orientations $\varphi_1$, $\varphi_2$ of the two
tubes within this unit cell satisfy the condition $\theta_1 =
\varphi_1 + \varphi_2 = 12^\circ$ and $\theta_2 = \varphi_1 -
\varphi_2 \approx {\pm}9.71^\circ$ within the range
$-18^\circ{\le}\theta<+18^\circ$.
 
Even though the energy barrier $\Delta{E}{\alt}0.5$~meV per atom for a
free rotation in a pair of tubes, shown in Fig.~\ref{Fig1}(b), appears
small, the barrier to rotate an entire tube, that is completely
straight and rigid, is unsurmountable. In the following, we postulate
that tube rotations in a rope are still possible in view of the
finite, albeit large value of the tube torsion constant. To determine
the torsional strain within an isolated, twisted $(10,10)$ tube, we
combined the LCAO method mentioned above with the recursion technique
\cite{SSCxRT}. This approach has been used successfully to describe
the dynamics of fullerene melting \cite{PRLxMLT}, the growth of
multiwall nanotubes \cite{PRLxLIP}, and the dynamics of a
``bucky-shuttle'' memory device \cite{PRLxNTM}. Our calculations,
shown in Fig.~\ref{Fig1}(d), suggest that the torsional energy is
harmonic up to a strain of $\Delta\varphi/\Delta
L{\approx}1^\circ/${\AA}. Within this harmonic regime, the torsional
energy {\em per atom} can be well represented by the expression
$\Delta{E}=\kappa(\Delta\varphi/\Delta L)^2$, with
$\kappa{\approx}2.58{\times}10^{-2}$~meV$\cdot$~rad$^2$/{\AA}$^2$.
Since the number of atoms in the tube is proportional to the total
tube length ${\Delta}L_{tot}$, the total torsional energy of the tube
is inversely proportional to the tube length for a given total twist
angle $\Delta\varphi_{tot}$ and hence becomes vanishingly small for a
long tube.
 
 \begin{figure}
    \centerline{
       \hspace*{0.12\columnwidth}
        {\raisebox{0.32\columnwidth}{\large\bf (a)}}
        \hspace*{-0.22\columnwidth}
        \epsfxsize=0.45\columnwidth
        \epsffile{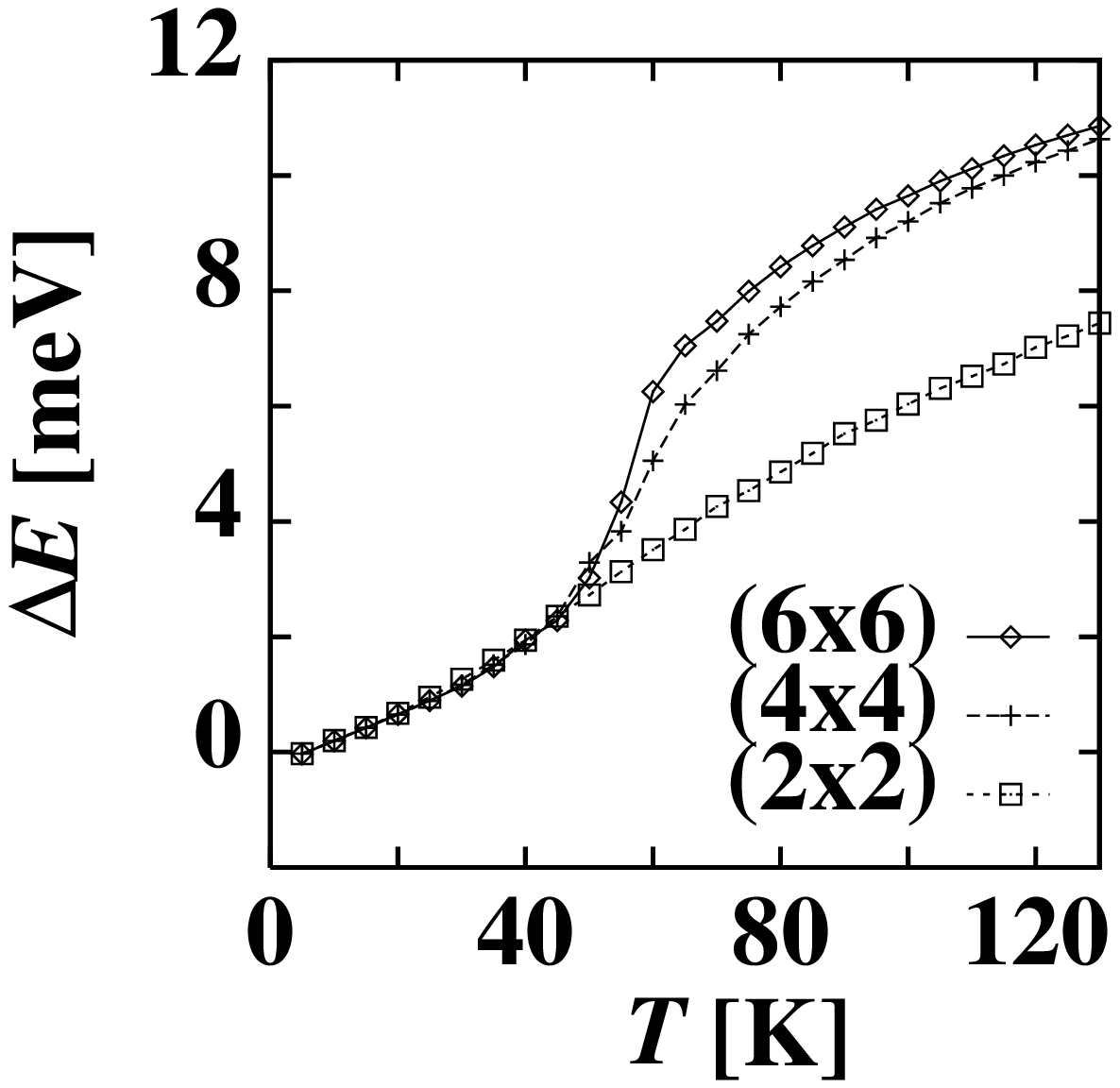}
        \hspace*{0.13\columnwidth}
        {\raisebox{0.32\columnwidth}{\large\bf (b)}}
        \hspace*{-0.22\columnwidth}
        \epsfxsize=0.45\columnwidth
        \epsffile{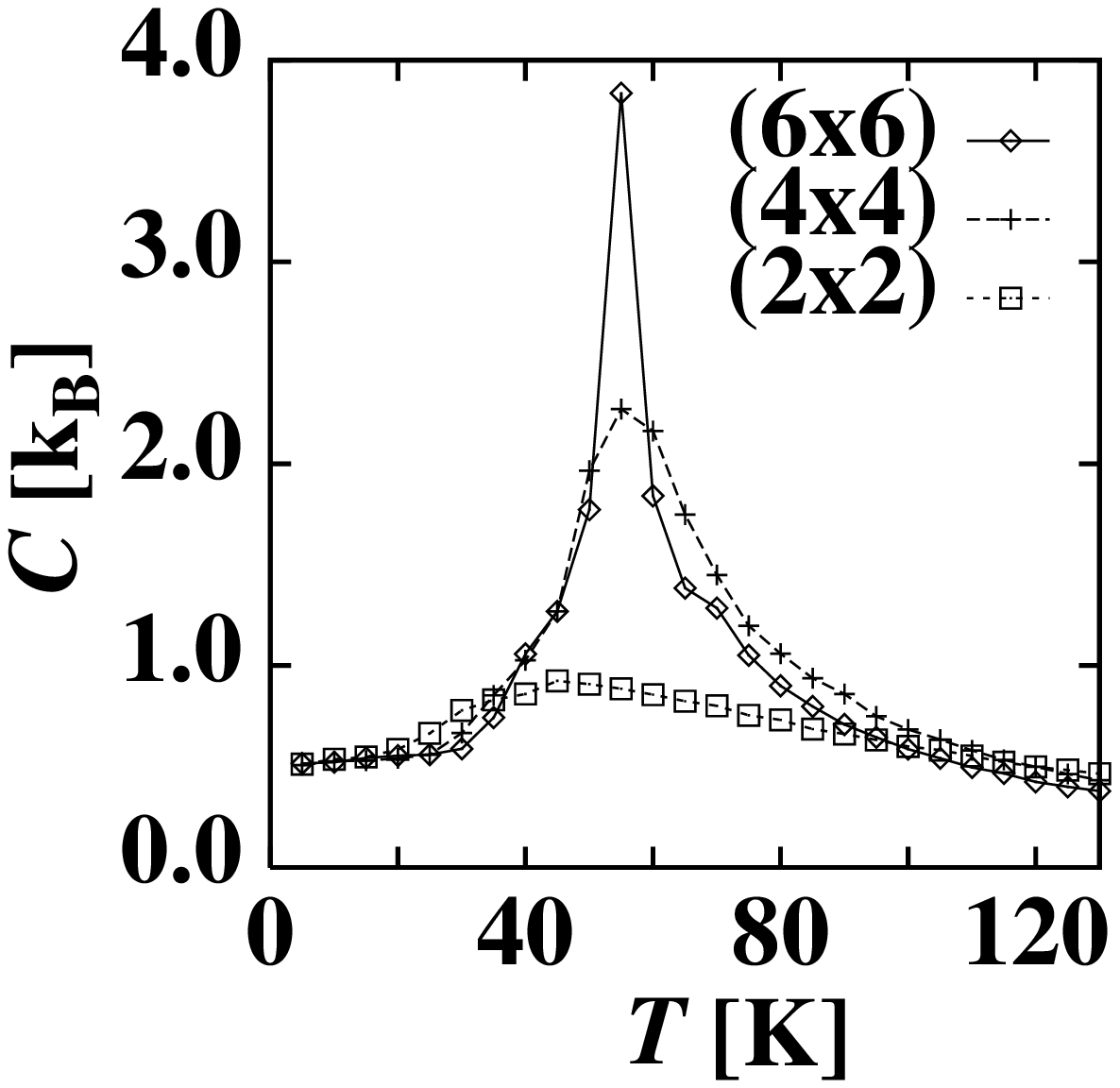}
    }
 \caption{
 Temperature dependence of
 (a) the total energy ${\Delta}E$ and
 (b) the specific heat $C$ {\em per degree of freedom} in model
 $(m{\times}m)$ superlattices of interacting (10,10) nanotubes with
 {\em zero torsional rigidity}. Each nanotube is modeled by a chain
 of torsionally decoupled, rigid segments (``disks'') containing
 twenty atoms, each representing one orientational degree of freedom.
 In this model system, orientational melting occurs at
 $T{\protect\agt}T_{OM}{\protect\approx}55$~K.
 }
 \label{Fig2}
 \end{figure}
 
To study the possibility of orientational melting in a nanotube rope,
we first consider an unrealistic model system of bundled $(10,10)$
tubes consisting of torsionally decoupled axial segments of 20 atoms,
resembling rigid ``rings'' or ``disks''. The interaction between two
adjacent disks in neighboring tubes is the value ${\Delta}E$ of
Fig.~\ref{Fig1}(b), multiplied by the number of atoms in the disk.
Absence of axial coupling makes this system equivalent to a
two-dimensional triangular lattice of disks with one (orientational)
degree of freedom per disk. Results of Monte Carlo simulations of
orientational melting in $(2{\times}2)$, $(4{\times}4)$, and
$(6{\times}6)$ superlattices with 4, 16, and 36 such tubes per unit
cell, respectively, are shown in Fig.~\ref{Fig2}. To ensure proper
phase space sampling even at low temperatures, each data point
represents an ensemble average taken over ${\agt}10^5$ Monte Carlo
steps {\em per degree of freedom}. Results for the temperature
dependence of the total energy per degree of freedom, shown in
Fig.~\ref{Fig2}(a), suggest that an orientational melting transition
should occur at $T_{OM}{\approx}55$~K. This transition becomes more
pronounced with increasing unit cell size in the superlattice. The
sharp peak in the temperature dependence of the corresponding specific
heat data, shown in Fig.~\ref{Fig2}(b), suggests this phase transition
to be of first order. At low and at high temperatures, the specific
heat approaches the classical value $C=0.5k_B$.
 
To address orientational melting in a realistic $(10,10)$ nanotube
rope, we add the proper torsional coupling between the disks,
according with our results in Fig.~\ref{Fig1}(d). Monte Carlo
simulations analogous to those described above showed no indication
that initially straight nanotubes would start to perform a
``ratcheting twisting motion'', corresponding to orientational melting
in the rope, up to $4,000$~K when individual nanotubes should
disintegrate structurally.
 
 \begin{figure}
    \centerline{
        \hspace*{0.12\columnwidth}
        {\raisebox{0.32\columnwidth}{\large\bf (a)}}
        \hspace*{-0.22\columnwidth}
        \epsfxsize=0.45\columnwidth
        \epsffile{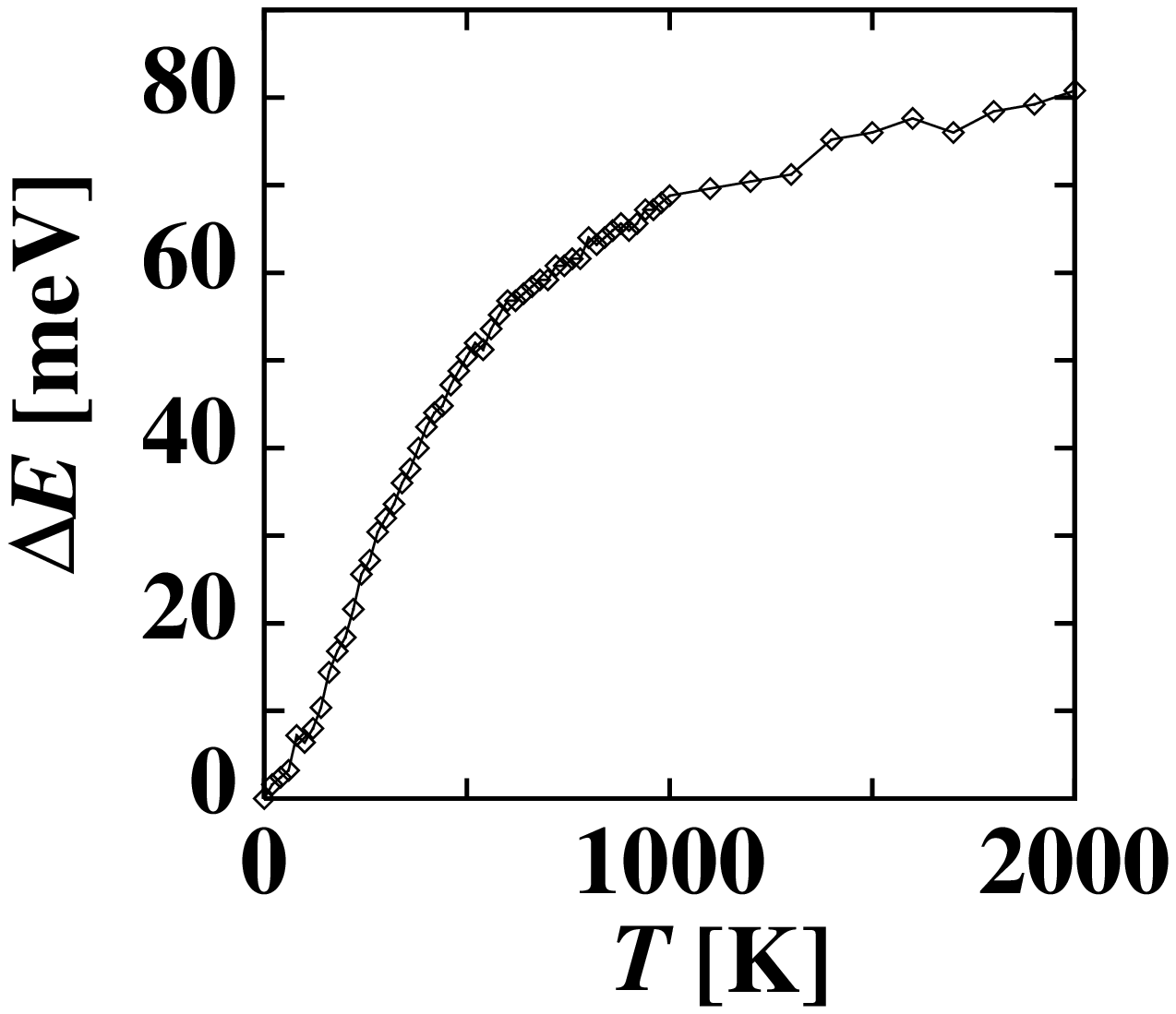}
        \hspace*{0.335\columnwidth}
        {\raisebox{0.32\columnwidth}{\large\bf (b)}}
        \hspace*{-0.42\columnwidth}
        \epsfxsize=0.45\columnwidth
        \epsffile{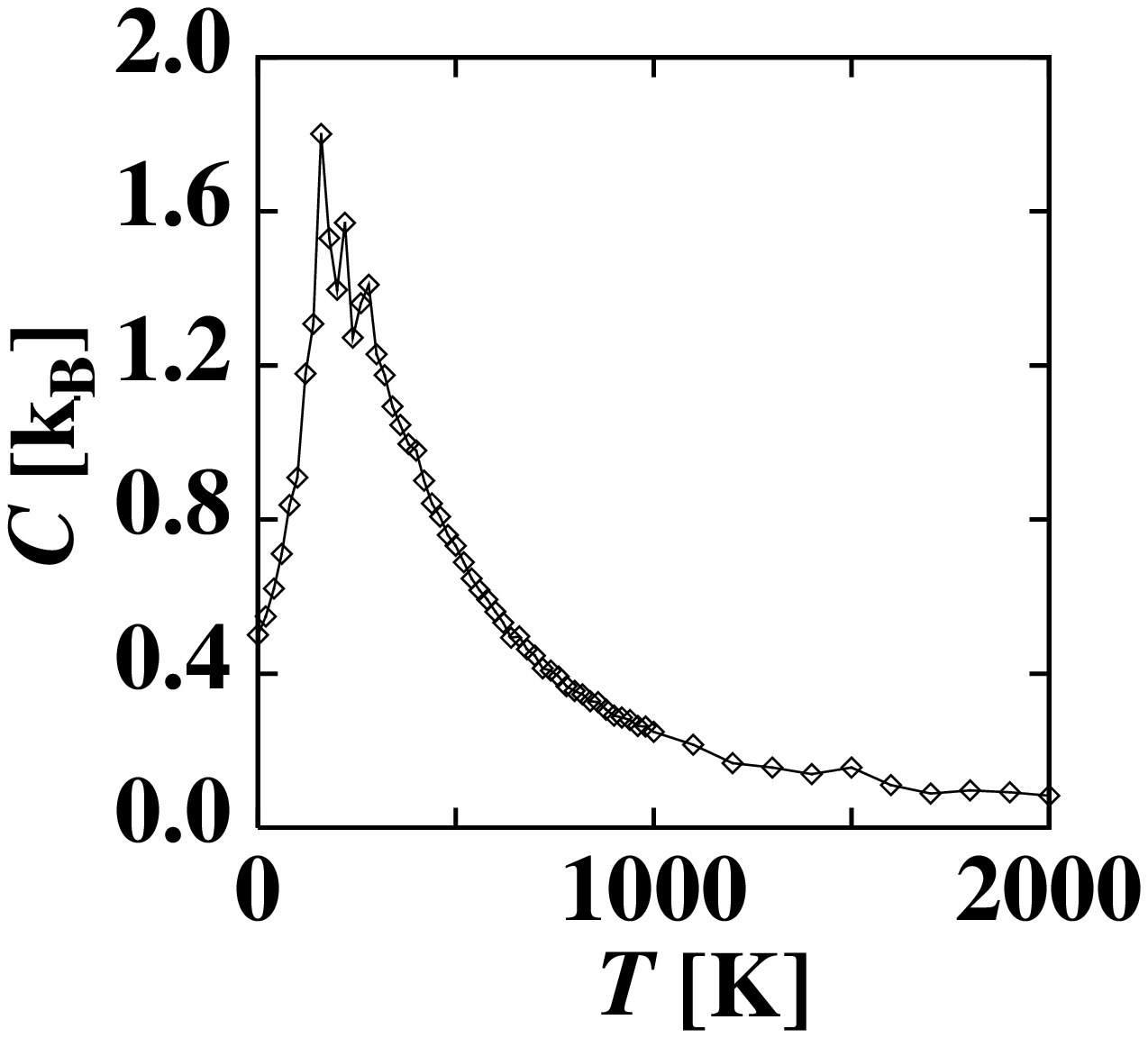}
    }
    \vspace*{0.03\columnwidth}
    \centerline{
        \hspace*{0.05\columnwidth}
        {\raisebox{0.55\columnwidth}{\large\bf (c)}}
        \hspace*{-0.1\columnwidth}
        \epsfxsize=0.45\columnwidth
        \epsffile{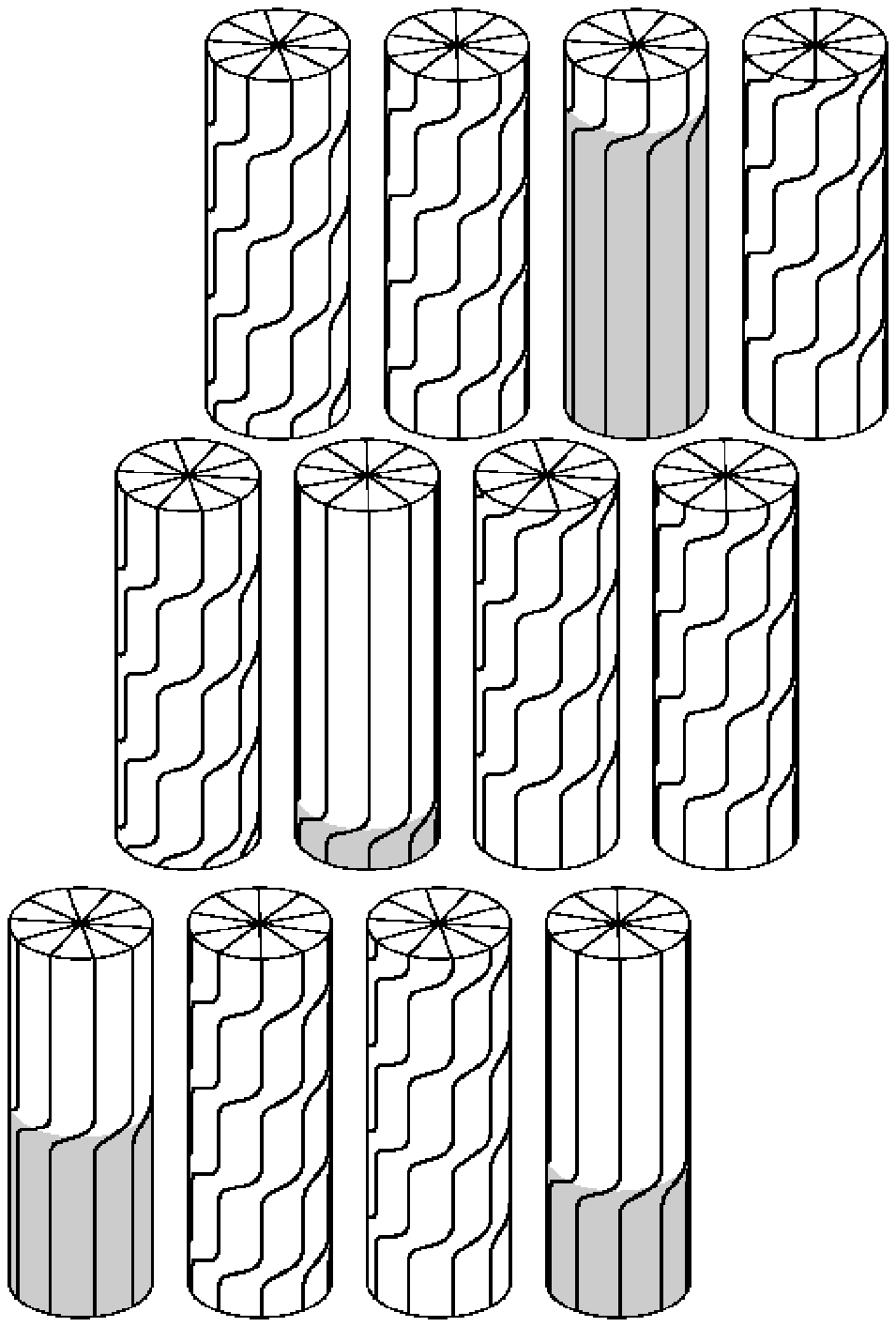}
        {\raisebox{0.32\columnwidth}{\large\bf $\Longrightarrow$}}
        \epsfxsize=0.45\columnwidth
        \epsffile{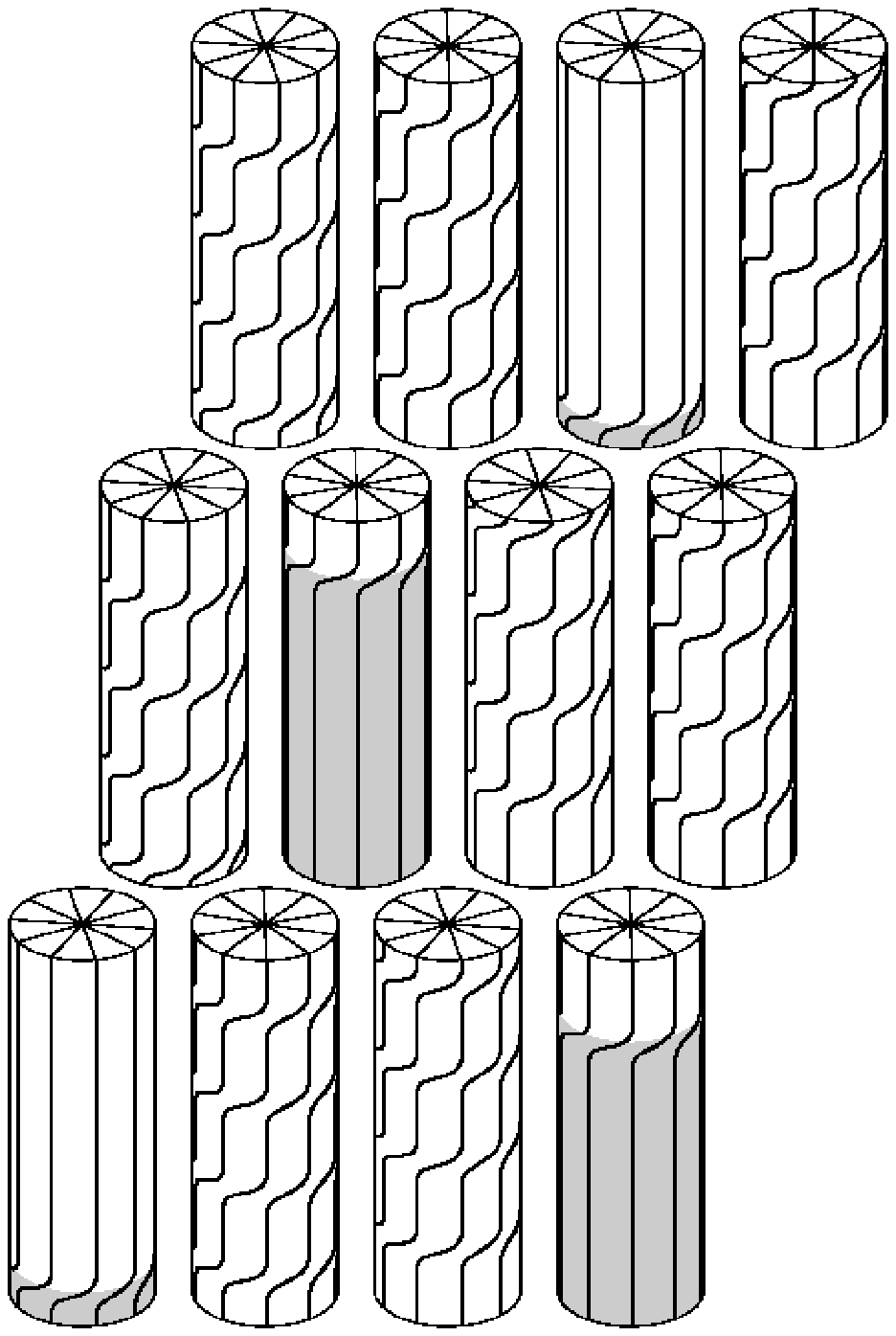}
    }
 \caption{
 Temperature dependence of (a) the total energy ${\Delta}E$ and
 (b) the specific heat $C$ {\em per degree of freedom} in a realistic
 rope consisting of interacting tubes of finite rigidity.
 (c) Illustration of the microscopic process of orientational melting
 in an exaggerated perspective. The two views depict the
 position of orientational dislocations at different points in time
 at $T=500$~K. The axial motion of the twistons, indicating
 orientational melting, is highlighted by the changing grey-shaded
 sections in four of the tubes.
 }
 \label{Fig3}
 \end{figure}

The key to the understanding of orientational melting is to consider
the energetics and dynamics of orientational dislocations in the
system. Consider two nanotubes which, under synthesis conditions, show
a total twist $\Delta\varphi_{tot}{\agt}36^\circ$ over the entire tube
length, at the negligible cost in total energy of ${\approx}0.1$~eV
for a $100~\mu$m long tube. As these tubes bundle up during annealing,
at least two orientationally aligned domains form within the tube
pair, separated by an orientational dislocation. Similar twists of up
to $1^\circ/$nm, associated with such orientational dislocations, have
been recently observed by electron diffraction in nanotube ropes
\cite{Loiseau}. Such solitons may move rather freely along the tube
axes, but cannot be annihilated if the paired tubes are infinitely
long, or if they bundle up to a double-torus \cite{Avouris}.
 
The equilibrium geometry of these frozen-in twistons is given by the
competition between the anisotropic inter-tube interactions, which
tend to align neighboring tubes, and the torsion rigidity that tends
to keep individual tubes straight. Their energetics and dynamics is
well represented by mapping the Frenkel-Kontorova model onto the
orientational degrees of freedom of a triangular lattice of chains. In
bundled $(10,10)$ nanotubes, the orientation ${\varphi}(z)$ of
individual tube layers near the dislocation is well reproduced by the
function
${\varphi}(z)=\varphi_0+
{\Delta}{\varphi}\left({1+\exp\left({(z-z_0)/w}\right)}\right)^{-1}$,
with the typical values ${\Delta}{\varphi}{\approx}36^\circ$ and
$w{\approx}30$~{\AA}. Such dislocations, which can be either left- or
right-handed, show a total twist of $36^\circ$ that extends over
${\approx}250$~{\AA} in the axial direction. The energy cost of
${\Delta}E{\approx}5$~eV to create such a twiston, given by the energy
difference between ropes containing one or no dislocation, is
relatively high. This explains why no twistons or left-/right-handed
twiston pairs were created in perfectly straight tubes even at high
temperatures. Since the twistons cannot be created easily, the
dynamics of the entire system is limited to a subspace of the
configurational space, where the number of twistons on each tube is
fixed. We notice that the axial motion of a twiston corresponds to a
finite tube rotation in that given segment, and that the energetics of
a rope containing twistons can be mapped onto a lattice gas model of
twistons moving along individual tubes. The small potential energy
barriers associated with an ``up''- and a ``down''-moving twiston
passing each other in adjacent tubes depend on the orientation of
the other surrounding tubes, and are evaluated by the total energy
expression above. We correlate the onset of orientational melting in
the rope with depinning and a completely free ``diffusion'' of
twistons within the tubes.
 
Results a Monte Carlo simulation for the orientational melting
transition in a $(10,10)$ nanotube rope containing twistons are
presented in Fig.~\ref{Fig3}. We consider a $(6{\times}6)$
superlattice of nanotubes with a fixed number of orientational
dislocations on each tube. The energy of this system is given by the
positions of these twistons within each periodically repeated unit
cell containing 36 tubes with 5,000 layer segments discretizing the
axial direction. This system with nominally 180,000 degrees of freedom
is mapped onto a lattice gas of twistons in the following way. Out of
the 36 tubes per unit cell, we select twelve non-adjacent tubes, each
containing a single twiston that can move axially. The remaining tubes
in the unit cell have eight such orientational dislocations frozen in.
Their positions are equally spaced over the tube axes, but axially
offset in adjacent tubes. The factual impossibility to change the
number of twistons, discussed above, is mimicked by assuming the same
handedness for all twistons.
 
The temperature dependence of the total energy of this classical
system, shown in Fig.~\ref{Fig3}(a), shows a slope that is initially
small close to $T=0$, then becomes large within the temperature
range $0$~K$<T{\alt}500$~K, and finally becomes small. The corresponding
specific heat data, shown in Fig.~\ref{Fig3}(b), begin with the
classical value $C=0.5k_B$ at low temperatures and peak at
$T_{OM}{\agt}160$~K, the orientational melting transition. The nature
of this transition is illustrated in Fig.~\ref{Fig3}(c). The two
snap-shots of the geometry, taken at $T=500$~K, indicate that above
$T{\approx}T_{OM}$, frozen-in twistons become depinned and diffuse
relatively freely along the tube axes, as indicated by the changing
gray-shaded sections in the tubes. Since these twistons are important
scattering centers for electrons, their depinning may significantly
affect the transport in these one-dimensional systems at
$T{\approx}T_{OM}$ that may play the role of $T^*$ in Ref.\ 
\onlinecite{Cond}.
 
 
In summary, using Monte Carlo simulations, we have investigated the
possibility of an orientational melting transition within a ``rope''
of $(10,10)$ carbon nanotubes. We postulate that during the
synthesis, as twisting nanotubes bundle up, orientational dislocations
or twistons arise from the competition between the anisotropic
inter-tube interactions, which tend to align neighboring tubes, and
the torsion rigidity that tends to keep individual tubes straight. We
have mapped the energetics of a rope containing such twistons onto a
lattice gas model and find that the onset of a free ``diffusion'' of
twistons, corresponding to orientational melting, should occur at
$T_{OM}{\agt}160$~K.
 
We acknowledge fruitful discussions with Marcel den Nijs and Jean
S.~Chung. This work was supported by the Office of Naval Research and
DARPA under Grant No. N00014-99-1-0252.
 



\begin{references}
\bibitem{cutoff} We use Fermi-Dirac expression
         $\left(1+\exp(\frac{d-d_c}{d_w})\right)^{-1}$, with
         $d_c=3.87$~{\AA} and $d_w=0.2$~{\AA}, as a
         smooth cutoff function of distance $d$ for both the
         hopping integrals and pairwise interactions.
\end{references}
\end{document}